\documentclass[prc,preprint,showpacs,doublespaced,article]{revtex4}

\usepackage{graphicx}
\usepackage{epstopdf}
\usepackage{color}
\usepackage{amsmath}
\usepackage{ulem}
\usepackage[T1]{fontenc}
\begin{document}

\title{{Studying VLOCV twin compact stars with binary mergers}}

\author{{ \bf Z. Sharifi\textsuperscript{1}\footnote{E-mail: sharifi@znu.ac.ir}, \bf M. Bigdeli\textsuperscript{1,2}\footnote{E-mail: m\underline\ \ bigdeli@znu.ac.ir}, \bf D. Alvarez-Castillo\textsuperscript{3,4}\footnote{E-mail: alvarez@theor.jinr.ru}}
}
\affiliation{\textsuperscript{1}Department of Physics, University of Zanjan, P.O. Box
45371-38791 Zanjan, Iran}
\affiliation{\textsuperscript{2}Research Institute for Astronomy and Astrophysics of Maragha, P.O. Box 55134-441 (RIAAM)-
Maragha, Iran}
\affiliation{\textsuperscript{3} {{H. Niewodnicza\'{n}„ski} Institute of Nuclear Physics,
Radzikowskiego 152, 31-342 Krak\'{o}w, Poland}}
\affiliation{\textsuperscript{4} Bogoliubov Laboratory for Theoretical Physics,
Joint Institute for Nuclear Research, 6 Joliot-Curie Street, 141980 Dubna, Russia}
\begin{abstract}
GW170817 has provided valuable constraints on the equations of state of merging binary neutron stars, which can be considered as the most probable candidate for the source of gravitational waves. On the other hand, these natural laboratories of extreme temperature and density may lead to the estimation of some exotic matter like {deconfined quark matter in their cores}. In this paper, we investigate the neutron star matter equation of state (EoS) with the lowest order constrained variational (LOCV) method considering the excluded volume effect (VLOCV) for nucleons {to} compute the tidal deformability of binary neutron star mergers (BNSMs). Within this approach, the size of nucleons makes the EoS {so} stiff that requires a phase transition in order to avoid causality violation. Therefore, this phase transition {may lead} to the appearance of the third family of compact stars {including} ``twin star'' configurations. {Our EoS models are confronted with observations from GW170817, GW190814, GW190425, and also NICER. We find out that regarding all these constraints, the EoS models having the {transition} pressure$\approx$30-100 MeV/fm$^{3}$ and the energy density discontinuity $\Delta\varepsilon$$\lesssim$300 MeV/fm$^{3}$ are preferable.}

\end{abstract}

\maketitle
\section{INTRODUCTION}
The study of nuclear matter equation of state (EoS) has long been {interesting} for both theoretical and experimental physicists. The EoS plays an important role in hydrodynamical models of nuclear collisions \cite{Rischke}. It is common to determine the nuclear matter properties by the energy per nucleon as a function of density \cite{costa}.
The most probable assumption for the nucleons in nuclear matter is that they can be considered as structureless particles with a bare mass $m_{N}$ caused by the strong interaction of quarks \cite{IJME}. The MIT bag model was invented to approve that nucleons are the composite systems of elementary particles \cite{MIT bag model}, which can be called extended objects.
One of the {simplest models that considers the finite} volume for nucleons is the Van der Waals (VDW) equation of state. The VDW model has the form as below \cite{PRC91}
\begin{eqnarray}
P(T,n)=\frac{nT}{1-bn}-an^{2},
\end{eqnarray}
in which $a>0$ and $b>0$ are VDW parameters describing attractive and repulsive interactions, respectively, and $n\equiv\frac{N}{V}$ is the {particle number density.}

A similar approach that takes into consideration this excluded volume resulting from the finite size of nucleons has been developed in \cite{Typel}. The resulting equation of state is rather stiff, its speed of sound growing unbound as the density increases to the point of causality breach where this approach is no longer valid. New degrees of freedom are expected in this high-density region, possibly in the form of deconfined quark matter entailing a phase transition in neutron star matter. {Transitions into quark matter may be crossover or first order, as considered in our manuscript. From a more fundamental point of view, the quark substructure of nucleons has to be considered. It leads to strong repulsion at high densities due to the action of the Pauli exclusion principle on the quark level provoking the deconfinement transition, see Ref. \cite{ben}. A recent study \cite{Cierniak:2020eyh} has shown that a special region exists for this quark matter model in the mass-radius diagram where hadronic configurations cannot be located. As a result, they postulate that a NICER measurement of the PSR J0740+6620 radius, 8.6-11.9 km, would be an indication of a quark matter core. {On the one hand}, data from heavy-ion collision experiments suggest deconfinement may occur in compact star interiors. On the other hand, binary neutron star mergers (BNSMs) are believed to have the capability of demonstrating the deconfinement of quarks at high densities and temperatures following the {event}~\cite{122}. {High density nuclear matter} in compact stars and {the appearance of} exotic states have been widely studied in many papers \cite{2,3,4,5}. As density increases, hadronic matter experiences two phase transitions: deconfining hadrons to quarks and gluons, and restoring chiral symmetry \cite{Erik2}. According to the fact that QCD is asymptotically free, the high-density and high-temperature phases consist of quarks and gluons, in which QCD symmetries are restored \cite{6}.} In the particular case that {the} phase transition is strong, fulfilling the so-called Seidov condition \cite{seidov}, a third branch in the mass-radius diagram of compact stars may appear. Moreover, a typical manifestation is mass twins stars: two stars of about the same mass but different radius lying in the second and third {branches} of the mass-radius diagram. This effect is merely due to their internal composition: the smaller star bears a quark matter core in contrast to the larger one, a pure hadronic star. A possible realization of twins stars, including realistic equations of state, can be found in Ref. \cite{ben}. In addition, the twins phenomenon has been extensively studied with Bayesian methods \cite{alvarez, alvarez2}, as well as in the framework of the physics of relativistic heavy-ion collisions \cite{alvarez3} and tidal deformabilities from gravitational wave signals \cite{alvarez4, alvarez5, Montana, chris}. A review of the astrophysical aspects of mass twins can be found in \cite{alvarez6}. On the contrary, different approaches for hybrid stars with a smooth transition at the interface do not lead to third branches in the mass-radius relations~{\cite{Ayriyan:2021prr}}. See, for instance, Ref.~\cite{124} for substitutional compounds phases or quarkyonic matter \cite{Zhaoo}. Interestingly, there exists a potential tension between the results of the compact star EoS based on nuclear calculations and multimessenger observations, see Ref. \cite{Biswas,Tan:2020ics}, for which third branch models with an early mass onset below the 1.4 M$_{\odot}$ provide a solution to this quandary.

In this work, we focus on the tidal parameters of BNSMs considering the volume for nucleons in neutron star matter, which is approximated by pure neutron matter {within the framework of the LOCV method.} In general, this model makes the EoS so stiff that the maximum mass of the neutron star increases substantially. In addition, {we consider an EoS \cite {Asadi} that features the three nucleon interaction (TNI) for the sake of comparison}. {At the same time}, the constant speed of sound (CSS) parametrization is utilized for the high-density region considering the first order phase transition from hadronic matter to quark matter. As a result, we revisit the four categories defined in Ref. \cite {Erik1} with our candidate EoS. In order to study the validity of the models considered in this work, we consider a set of constraints from multi-messenger observations, i.e., x-ray observations of PSR J0030+0451, and gravitational wave detections of mergers, events GW170817, GW190814, and GW190425. The latter event corresponds to the second observation of a gravitational-wave signal indicating the coalescence of a BNS system \cite{1908425}. {It has been estimated} that the masses of components range from 1.12 to 2.52 M$_\odot$ for the high-spin prior (1.46-1.87 M$_\odot$ for the low-spin prior). It is to be noted that the chirp mass 1.44$_{-0.02}^{+0.02}$ M$_\odot$ and the total mass 3.4$_{-0.1}^{+0.3}$ M$_\odot$ of this binary are larger than any known BNS system. Recently, Tatsuya Narikawa \textit{et al.} reanalyzed the BNSMs GW170817 and GW190425 employing a numerical-relativity calibrated waveform model \cite{reanalysis}. {Unluckily}, the binary system in the GW190425 event is massive and little information can be obtained on its tidal deformability. For instance, they reported that this event restricts $\widetilde{\Lambda}$ to be $\leq$700 using TF2+\_ PNTidal model for the low-spin prior.

This article is organized as follows. Section II includes the models and formalisms, namely the LOCV approach considering the excluded volume effect in order to investigate the equation of state of neutron star matter. The formalism for the bag model is also presented in this section for taking into account the changes of the structure of nucleons inside the compressed nuclear matter. Section III refers to the CSS parametrization, classifications of twin stars, and the results provided for the mass-radius relation within our EoS models. {BNSMs tidal deformability results for our two candidate approaches are presented in Sec. IV. Finally, Sec. V includes a brief summary of the results and concluding remarks.}

\section{FORMALISM}
We consider neutron matter as an infinite system of strongly interacting $A$ neutrons with finite size.  The proper volume defined as,
\begin{eqnarray}\label{tener}
          b=\frac{16}{3}\pi r^3,
 \end{eqnarray}
where $r$ is the hard-sphere radius of neutrons. The number density of the system is $\rho=A/V$. 
The energy per particle and pressure of such a system can be obtained from the VLOCV model as
\begin{eqnarray}
         E(\rho)&=&E^*_{nuc}(\rho^*),\\ \nonumber
         p_H(\rho)&=&\rho^2 \frac{\partial E(\rho)}{\partial \rho}={\rho^*}^2 \frac{\partial E^*(\rho^*)}{\partial \rho^*}=p^*_H(\rho^*),
         \end{eqnarray}
where $\rho*={\rho}/{(1-b\rho)}$ and $E^*_{nuc}$ is the energy per nucleon which is calculated by using the LOCV method as follows.


We adopt a trial many-body wave function of the form, $\psi=\cal{F}\phi$,
where $\phi$ is the uncorrelated ground state wave function of $A$
independent nucleons, ${\cal F}={\cal S}\prod _{i>j}f(ij)$ is a Jastrow form of an appropriate
$A$-body correlation operator {, and ${\cal S}$ is a symmetrizing operator.} 
 %
Now, we consider the cluster expansion of the energy, in terms of correlation function, $f$, and its derivatives, functional up
to the two-body term \cite{clark},
 \begin{eqnarray}\label{ten1}
           E^*_{nuc}([f])=\frac{1}{A}\frac{\langle\psi|H|\psi\rangle}
           {\langle\psi|\psi\rangle}=E _{1}+E_{2}\cdot
 \end{eqnarray}
$E_1$ is the one-body term,
\begin{eqnarray}\label{ener1}
               E_{1}= {\sum _{k\leq{k_n^F}}
               \frac{\hbar^{2}{k^2}}{2m_n}},
 \end{eqnarray}
where
$k_n^F=3\pi^2\rho^*$ is the fermi momentum.
%
%

In Eq. (\ref{ten1}), the two-body energy $E_2$ can be written as
\begin{eqnarray}
E_{2}=\frac{1}{2A}\sum_{ij} \langle ij\left|-{\hbar^{2}}/{2m}[f(12),[\nabla_{12}^{2},f(12)]]+f(12)V(12)f(12)\right|ij-ji\rangle ,
\end{eqnarray}
which is as a function of two-body correlation operator $f(12)$~\cite{OBI3}, and two-body
potential $V(12)$. In our calculations, we use the two-body
potential AV18 \cite {wiringa} with and without TNI \cite{Asadi}. {Applying the contribution of TNI \cite{Lagaris} to this potential causes the saturation properties of symmetry energy to be in good agreement with the experiment,
\begin{eqnarray}
E=E(AV_{18}+TNR)+TNA.
\end{eqnarray}
Here, $TNA=\gamma_{2}\rho^{2}exp(-\gamma_{3}\rho)(3-2\beta^{2})$ refers to the three nucleon attractive part of TNI. TNR is the repulsive part given by multiplying the tensor function by a factor $exp(-\gamma_{1}\rho)$. The $\gamma_{i}$ parameters are chosen so that the saturation properties of symmetric nuclear matter can be fulfilled (See Ref. \cite{Asadi} for the details).}
From the functional minimization of the two-body cluster energy with
respect to the variations in the correlation functions but subject to the normalization constraint,
\begin{eqnarray}
        \frac{1}{A}\sum_{ij}\langle ij\left| \left[ 1-\frac{9}{2}\left( \frac{J_{J}^{2}(k_\tau^F r)}{k_\tau^F r}\right) ^{2}\right] ^{-1}-f^{2}(12)\right| ij\rangle
        _{a}=0,
 \end{eqnarray} we get a set of coupled and uncoupled Euler-Lagrange differential equations
\cite{OBI3,BM98}. By solving these differential equations, we can
obtain correlation functions to compute the two-body energy term.

{The properties of nucleons, e.g. mass and radius, inside the compressed nuclear matter can be altered. However, the change in the nucleon invariant bare mass m$_{N}$ is negligible at the saturation density $\rho_{0}$ in comparison with its value in vacuum according to the deep inelastic phenomenology (See Ref. \cite{IJME} and the references therein). Thus, the assumption of constant nucleon mass in nuclear matter is taken into account in our work.
The bag model is one of the simplest models to describe the nucleons as the particles with structure. In this model, nucleons have the spherical volume of $\Omega_{N}$ as the systems with internal components of three quarks in the lowest state, which have the energy} \cite{MIT bag model}
\begin{eqnarray}
E_{Bag}^{0}=\frac{3\omega_{0}-Z_{0}}{r_{0}}+\frac{4\pi}{3}B(\rho_{0})r_{0}^{3}.
\end{eqnarray}
\begin{figure}[b]
\centerline{\includegraphics[scale=0.6]{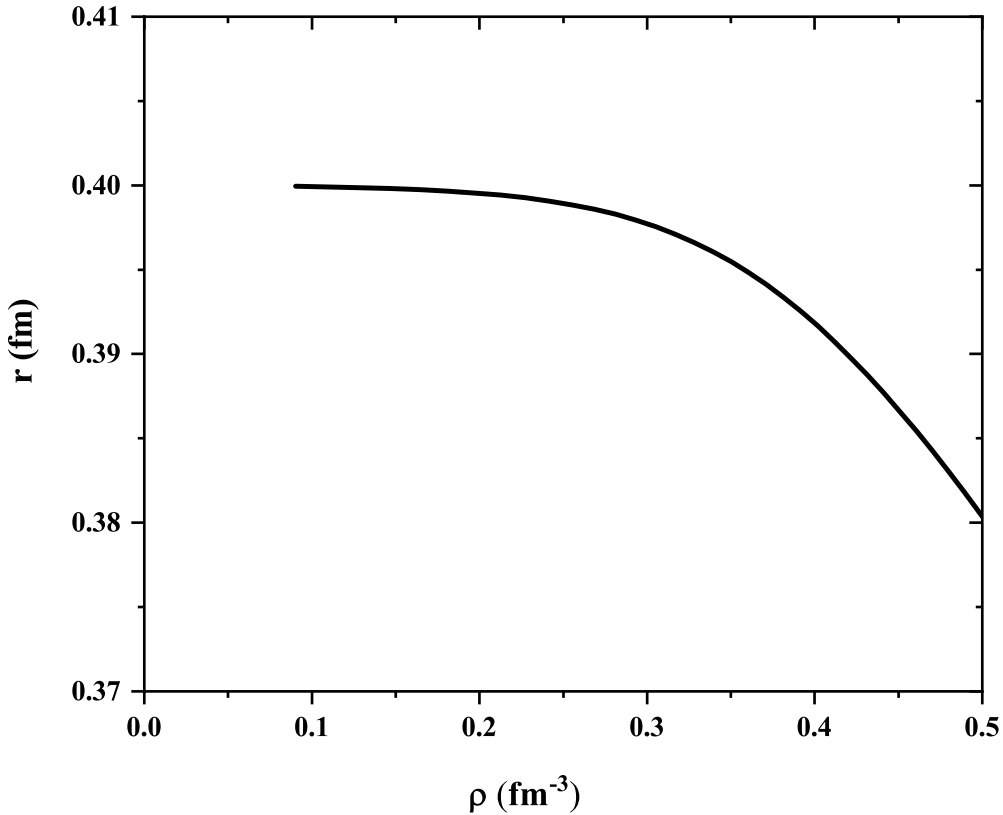}}
\caption{\small Radius of nucleon as a function of density for r$_ 0=$0.4 fm.}\label{ro}
\end{figure}
\begin{figure}[b]
\centerline{\includegraphics[scale=0.6]{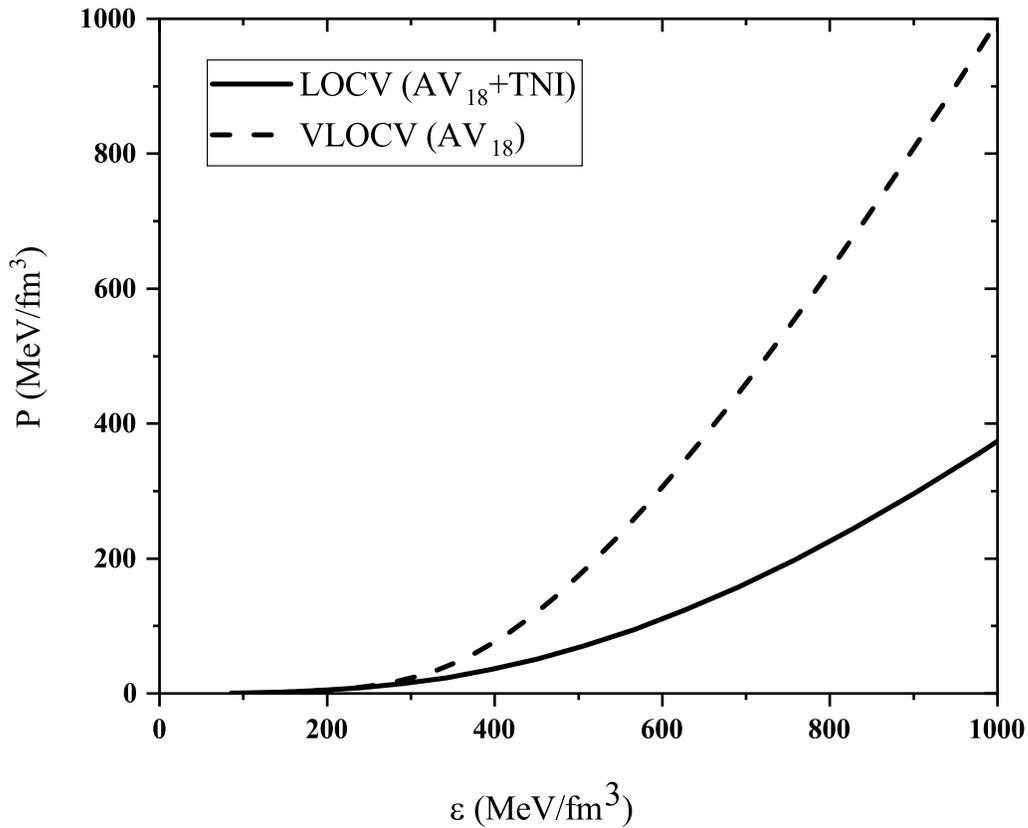}  }
\caption{\small Candidate equations of state with AV$_{18}$ potential considering the impact of nucleon size and TNI.}\label{pe}
\end{figure}Here, $r_{0}$ is the radius of nucleon at saturation density, $\omega_{0}$ and $Z_{0}$ are phenomenological constants, and $B(\rho)$ is the bag constant (for the details, see Refs. \cite{quark meson,Gue,guch}). {The following condition}
\begin{eqnarray}
p_{B}=-(\frac{\partial E_{Bag}^{0}}{\partial \Omega_{N}})_{surface}=0,
\end{eqnarray} 
for the pressure inside the bag $p_{B}$ generated by the partons yields the relation between $r_{0}$ and B{,
\begin{eqnarray}
r_{0}=\left[\frac{3\omega_{0}-Z_{0}}{4\pi B(\rho_{0})}\right]^{1/4}\leftrightarrow B(\rho_{0})=\frac{3\omega_{0}-Z_{0}}{4\pi r_{0}^{4}}
\end{eqnarray}
Replacing E$_{Bag}^{0}$ by the nucleon mass m$_{N}$ at saturation density and applying Eq. (11) to Eq. (9) leads to the following equation,
{\begin{eqnarray}
 3\omega_{0}-Z_{0}=\frac{3}{4}m_{N}r_{0}.
\end{eqnarray}}}{Assuming that in a compressed medium, $p_{B}$ is equal on the bag surface to the nuclear pressure $p_{H}$, generated by the elastic collisions of other hadrons, the density dependent radius of the nucleon is given by}
\begin{eqnarray}
p_{H}=p_{B} \rightarrow r({\rho})=\left[\frac{3\omega_{0}-Z_{0}}{4\pi (B(\rho)+p_{H}(\rho))}\right]^{1/4}{\leftrightarrow B(\rho)=\frac{3\omega_{0}-Z_{0}}{4\pi r^{4}(\rho)}-p_{H}(\rho)}.
\end{eqnarray}
{A mass m($\rho$) in nuclear medium can be obtained from Eq. (9) \cite{IJME, nuclear}:
\begin{eqnarray}
m(\rho)=\frac{3\omega_{0}-Z_{0}}{r(\rho)}+\frac{4\pi}{3}B(\rho)r^{3}(\rho).
\end{eqnarray}
Using Eqs. (12, 13), the above equation takes the form as below
\begin{eqnarray}
m(\rho)=m_{N}\frac{r_{0}}{r(\rho)}-p_{H}\Omega_{N}.
\end{eqnarray}
The assumption of constant nucleon mass $m(\rho)=m_{N}$ in nuclear medium results in
\begin{eqnarray}\label{cm}
m_{N}r(\rho)+p_{H}\frac{4\pi}{3}r^{4}(\rho)=m_{N}r_{0}.
\end{eqnarray}
}
\begin{table*}[t]
 \caption{{\label{max} The maximum mass (M$_{\odot}$) of neutron star
 obtained for the candidate EoSs.}}

\begin{tabular}{cccc}
   \hline
    \hline
   EoS & LOCV (AV$_{18}$)~\cite{Elyasi}\  \ &LOCV ({AV$_{18}$}+TNI)~\cite{Asadi}\  \ & VLOCV (AV$_{18}$) \\
  \hline
    M$_{max}$        &1.623\ &2.11&2.19\\

    \hline
    \hline

   \end{tabular}
   \end{table*}
Obtaining the nuclear pressure $p_H$ by the VLOCV method, r($\rho$) can be solved in this equation. Therefore, we have plotted the radius of nucleon as a function of density within our approach for r$_0=$0.4 fm {in Fig. \ref{ro}.} One can observe that the radius of nucleon remains approximately constant for the densities of $\simeq$(0.1-0.25) fm$^{-3}$, and then its value reduces slightly.

 Figure \ref{pe} displays our candidate EoSs for the hadronic phase in order to examine the tidal deformabilities of twin star solutions. It can be deduced that the EoS obtained by our new approach leads to a great stiffness. The results of maximum mass obtained for three different EoSs with AV18 potential, which manifests the interaction of two nucleons, have been demonstrated in Table \ref{max}. It is obvious that {including} TNI fulfills the maximum mass constraint for neutron stars, while the VLOCV approach makes the EoS very stiff so that the causality condition is violated at the density of 0.45 $fm^{-3}$. The value calculated for the maximum mass before this violation is 2.19 M$_{\odot}$. {On the contrary}, the EoS without {TNI plus finite size effects} is so soft that the maximum mass constraint cannot be {fulfilled}.

\section{HYBRID EQUATION OF STATE WITH FIRST ORDER PHASE TRANSITION}
Deconfinement of hadrons is probable when the density {inside compact stars} increases, and as a result, the quark matter core arises. As we previously mentioned, we use our calculated EoSs applying VLOCV (AV$_{18}$) and LOCV {(AV$_{18}$+}TNI) approaches for the hadronic phase, while {employing} the widely used parametrized CSS equation of state presented by Alford for the high-density quark phase \cite{Alford}. This EoS, which has three control parameters, {is parametrized as shown below} for a {fixed} hadronic EoS,
\begin{equation}
    \varepsilon=
\begin{cases}
    \varepsilon(p) \quad \quad \quad \quad \quad \quad \quad \quad \quad \quad \quad \quad \quad\quad    p<p_{trans}\\
    \varepsilon(p_{trans})+\Delta\varepsilon+c_{QM}^{-2}(p-p_{trans})  \quad \quad      p>p_{trans}\\
\end{cases}
\end{equation}
where $\varepsilon$, $p$, $p_{trans}$, and $\Delta\varepsilon$ are energy density, pressure, {transition} pressure, and energy density discontinuity at the transition, respectively. It is to be noted that the speed of sound in the core of stars $c_{QM}$ is assumed to be 1.

Moreover, for large discontinuity in energy density at the transition, the star undergoes instability when central pressure equals {transition} pressure. {This will happen whenever the value of $\Delta\varepsilon$ is equal or greater than the so called \textit{Seidov limit} \cite{seidov}:}
\begin{equation}
\frac{\Delta\varepsilon_{crit}}{\varepsilon_{trans}}=\frac{1}{2}+\frac{3}{2}\frac{p_{trans}}{\varepsilon_{trans}},
\end{equation}
where $\Delta\varepsilon_{crit}$ is {the} threshold value, while $\varepsilon_{trans}$ and $p_{trans}$ are energy density and pressure at the phase transition{, respectively \cite{Erik1}. All the EoSs in this work fulfill the above equality, thus leading to third branches in the mass-radius relation.}
\subsection{Categories of twin stars}
Twin stars, consisting of neutron stars with similar {masses} but different sizes, originate from {a EoS with a strong} first order phase transition \cite{glendening, zacchi, sch}. As {introduced and} defined in \cite{Erik1, Erik2}, it {is of} a great advantage to categorize twin star solutions into four groups. Therefore, a brief definition of these classes of twin stars is provided {here}. {We start by} noting that M$_{1}$ and M$_{2}$ refer to {their respective} maximum masses of hadronic and twin star branches, related to nonrotating {compact stars}. In fact, the structural parameters of the neutron star can be obtained by solving the Tolman-Oppenheimer-Volkoff (TOV) equations \cite{Tolman,Oppenheimer},
\begin{eqnarray}
&&\frac{dP}{dr}=-\frac{\varepsilon m}{r^2}(1+\frac{P}{\varepsilon })(1+\frac{4\pi Pr^3}{m})(1-\frac{2m}{r})^{-1}\nonumber\\&&
\frac{dm}{dr}=4\pi r^2 \varepsilon.
  \end{eqnarray}

\begin{figure}[t]
\centerline{\includegraphics[scale=0.6]{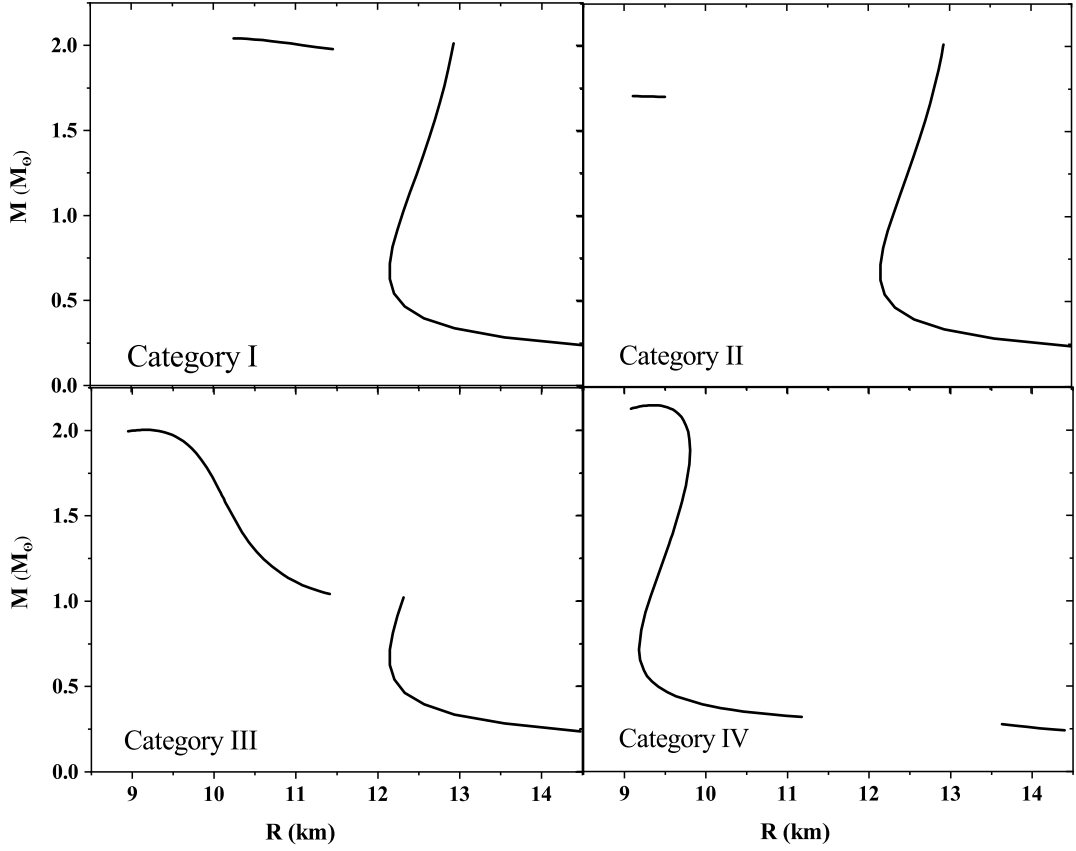}  }
\caption{\small The schematic trend of mass-radius relation for the classes of twin stars, I-IV, defined in the text.}\label{four}
\end{figure}

Samples of these categories are displayed in Fig. \ref{four} and defined as follows \cite{Erik1}:
\subparagraph{I:} Both M$_{1}$ and M$_{2}$ exceed 2 M$_{\odot}$ due to the high values of $p_{trans}$. Moreover, the twin branch is approximately flat.
\subparagraph{II:} The conditions of M$_{1}\geq 2$ M$_{\odot}$ and M$_{2}<$ 2 M$_{\odot}$ are satisfied, which manifests the high values of $p_{trans}$ {found only in massive compact stars.}
\subparagraph{III:} The hadronic branch has the property of $1 M_{\odot} \leq M_{1}\leq 2 M_{\odot}$, however, M$_{2}$ exceeds 2 M$_{\odot}$. Therefore, $p_{trans}$ is lower compared to the previous classes and the twin branch becomes steeper.
\subparagraph{IV:} The condition for M$_{2}$ is the same as category III, while M$_{1}$ is below 1 M$_{\odot}$. {Consequently,} the twin branch has the steepest slope in this category.
\begin{table*}
 \caption{{\label{twinn} The categories of twin stars defined by the values of maximum mass associated with the hadronic and hybrid star branches applying the EoS discussed in the text. {Columns labeled \textit{low} and \textit{high} correspond to the lowest and highest values of the CSS EoS parameters for each category.} $p_{trans}$ and $\Delta\varepsilon$ {values} are given in units of MeV/fm$^{3}$.}}
 \begin{tabular}{ccccc}
   \hline
    \hline
   VLOCV (AV$_{18}$)&Low $p_{trans}$&  High $p_{trans}$ &Low $\Delta\varepsilon$&  High $\Delta\varepsilon$\\
  \hline
  CI        &108&130&380&430\\
  
   CII       &108 &127&420&700\\
   
  CIII   &31 &107&250&415\\

   C IV   &5&30&250&430\\
 
 LOCV ({AV$_{18}$+}TNI)\\
   
  CIII   &40 &41&250&273\\
   C IV   &4&39&250&445\\
    \hline
     \hline

   \end{tabular}
   \end{table*}

\begin{figure}[!ht]
\centerline{\includegraphics[scale=0.6]{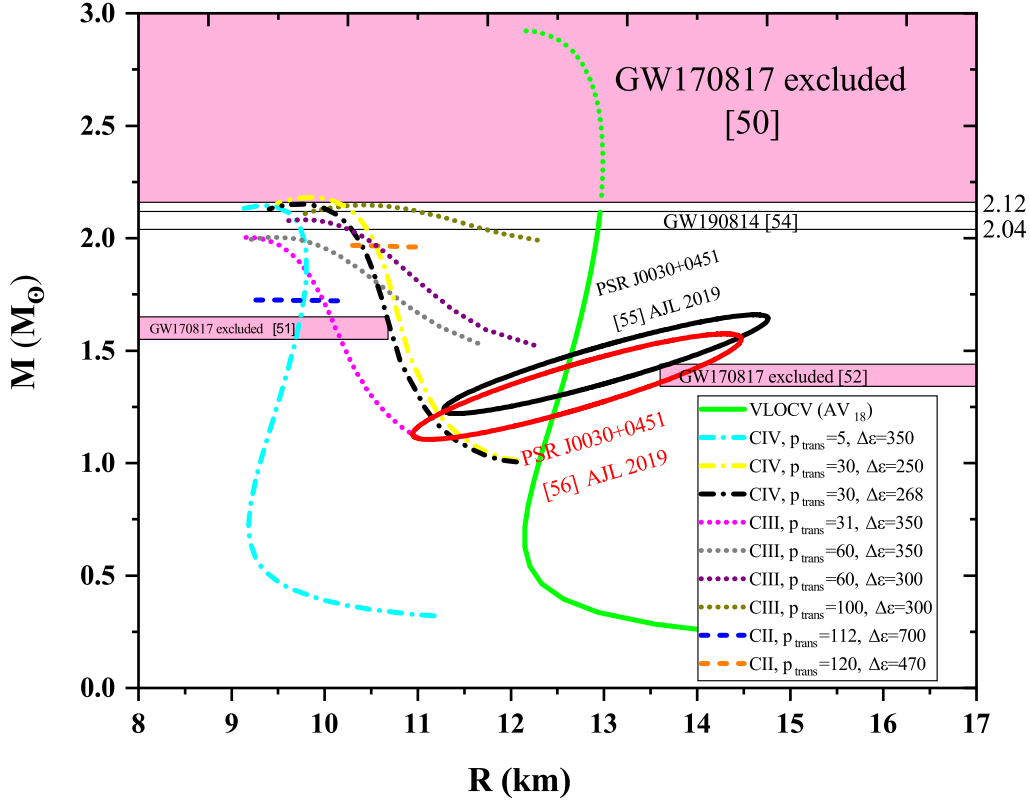}  }
\caption{\small The mass-radius relation of VLOCV (AV$_{18}$) EoS with categories II, III, and IV. {It should be noted that the values of $p_{trans}$ and $\Delta\varepsilon$ are given in units of MeV/fm$^{3}$.}}\label{WM}
\end{figure}
\begin{figure}[!ht]
\centerline{\includegraphics[scale=0.6]{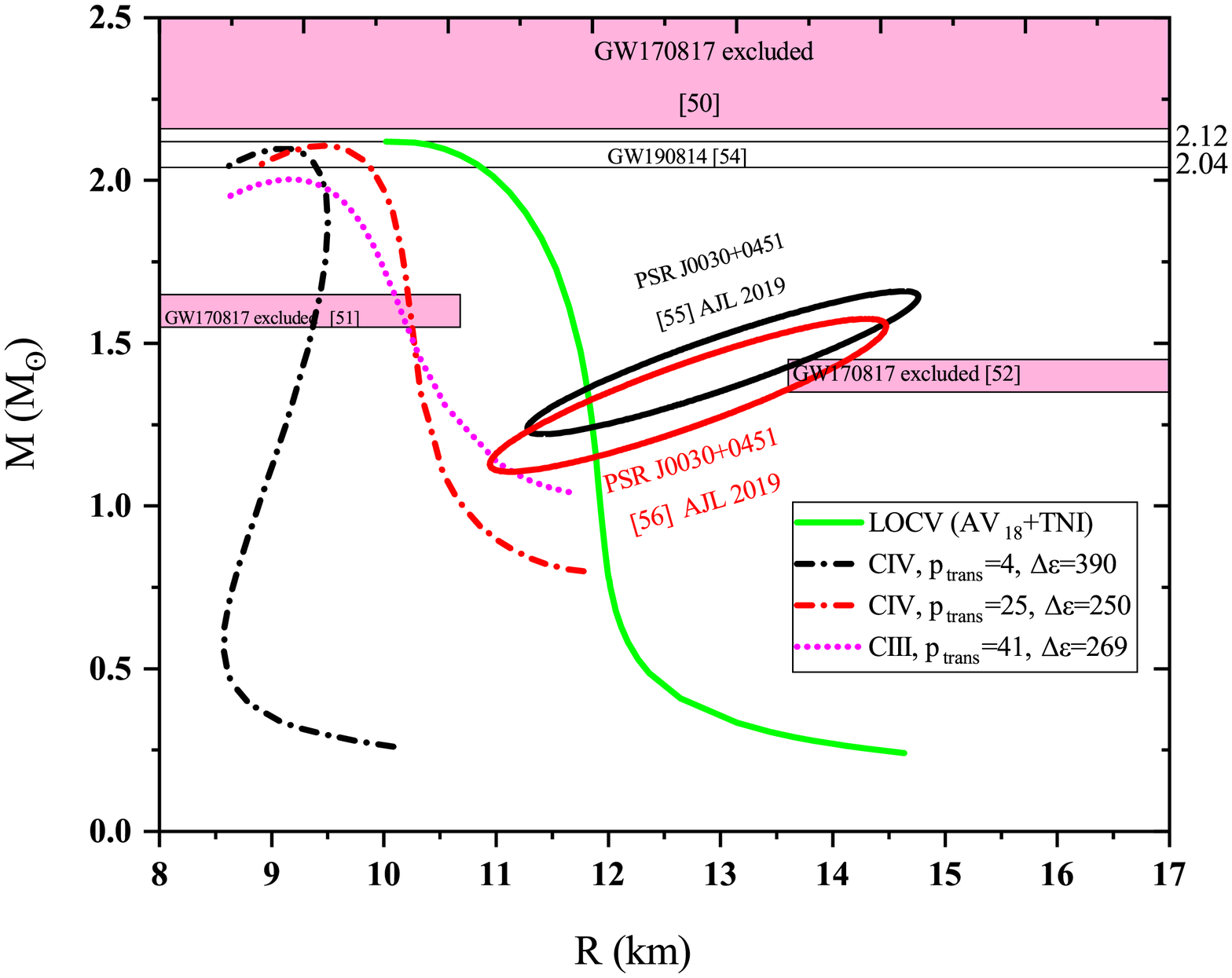}  }
\caption{\small The mass-radius relation of LOCV ({AV$_{18}$+}TNI) EoS with categories III and IV. {The values of $p_{trans}$ and $\Delta\varepsilon$ are given in units of MeV/fm$^{3}$.}}\label{pM}
\end{figure}
  {Table \ref{twinn} displays the high density parameters $p_{trans}$ and $\Delta\varepsilon$ for the two EoS approaches in our study.} We have obtained the {lowest} and {highest} limits of these parameters for each category, which indeed satisfies the definitions of their classification by mass. {Category I would be only of interest for M$_{total}\geq4$ M$_{\odot}$ by definition. Moreover, since the estimated mass values of the components of GW170817 are below 2 M$_{\odot}$, the corresponding} {stars} {described by this category must be pure hadronic configurations. Category II, whose second branch is flat, implies high $p_{trans}$, (108-127) MeV/fm$^{3}$, with the maximum mass of above 2 $M_{\odot}$ in the hadronic branch. The values of $p_{trans}$ within the range (31-107) MeV/fm$^{3}$ satisfy the definition of category III and lead to a steeper mass increase compared to categories I and II. The lowest values of $p_{trans}$ (5-30) MeV/fm$^{3}$ generate category IV, which results from the definition that the maximum mass in the hadronic branch} {must be} {below 1 $M_{\odot}$ and $M_{max}$ in the hybrid branch} {must be} { higher than 2 $M_{\odot}$.} {It is worth mentioning that} the conditions for category I and category II have not been produced for the EoS with LOCV {(AV$_{18}$+}TNI).

{The corresponding mass-radius diagrams for the two candidate EoS approaches are shown in Figs. \ref{WM} and \ref{pM}, respectively.} {These plots are typical examples of different $p_{trans}$ and $\Delta\varepsilon$ values related to every possible category, as presented in Table \ref{twinn}. Due to the fact that category I dissatisfies the constraint of the total mass of BNSM in GW170817, $2.73_{-0.01}^{+0.04}$ $M_{\odot}$, its compact star sequence is displayed in neither figure. Furthermore, some constraints obtained from recent observations are {also} shown in these figures. Utilizing the observation of the GW170817 event, modeling of GRB 170817A, and the quasiuniversal relations, Rezzolla \textit{et al.} set limits for the maximum mass of nonrotating stars to lie in the range 2.01$_{-0.04}^{+0.04}\leq M_{TOV}/M_{\odot}\lesssim 2.16_{-0.15}^{+0.17}$ \cite{Rezzollaa}. {Additionally}, the radius of the stellar structure of 1.6 M$_{\odot}$ nonrotating NSs   {must be} larger than 10.68$_{-0.04}^{+0.15}$ km, which is obtained by the binary mass measurement of GW170817 and the assumption of delayed collapse in this event \cite{Bauswein}. Moreover, Annala \textit{et al.} demonstrated that the maximum radius of a 1.4 M$_{\odot}$ NS is 13.6 km \cite{Annala}. According to the most recent detection GW190814 \cite{190814}, Most \textit{et al.} estimated a lower boundary on the maximum mass of nonrotating neutron stars: M$_{TOV}>2.08^{+0.04}_{-0.04}M_{\odot}$ \cite{Mostt}. They achieved this result by stating that the secondary in this event was either a black hole resulting from the rapidly rotating NS collapse or a stable rapidly rotating NS. The recent x-ray observation of the periodic signal of the object PSR J0030+0451 by NICER has reported either M=1.44$_{-0.14}^{+0.15}$ M$_{\odot}$ with R = 13.02$_{-1.06}^{+1.24}$ km \cite{miller} or M =1.34$_{-0.16}^{+0.15}$ M$_{\odot}$ with R = 12.71$_{-1.19}^{+1.14}$ km \cite{Riely}.} {In Fig. \ref{WM}, one can observe that all the displayed hybrid stars generated by the VLOCV model satisfy the maximum mass constraint presented by Rezzolla \textit{et al.} \cite{Rezzollaa} except the one related to the category IV with $p_{trans}$=30 MeV/fm$^{3}$ and $\Delta\varepsilon$=250 MeV/fm$^{3}$. {It is also apparent in this figure that low values of $p_{trans}$ for categories III  and IV, 5 and 31 MeV/fm$^{3}$, respectively, together with particular values of $\Delta\varepsilon$ result in the hybrid stars that do not fulfill the constraint presented by Bauswein \textit{et al.} \cite{Bauswein}.} {Furthermore}, it can be seen that all the EoS models shown in this figure satisfy the constraint set by Annala \textit{et al.} \cite{Annala}. In general, the {transition} pressure in each mentioned category should increase in order to meet the constraint set for the radius of 1.6 M$_{\odot}$ NS structure. For instance, changing the value of $p_{trans}$ from 5 MeV/fm$^{3}$ to 30 MeV/fm$^{3}$ in category IV leads to a better result in accordance with this criterion. The maximum mass boundary obtained by GW170817 puts another constraint on our EoS models of category IV, for which the {low} value of $\Delta\varepsilon$ reported in Table \ref{twinn} should rise from 250 MeV/fm$^{3}$ to 268 MeV/fm$^{3}$. Furthermore, the maximum value of this parameter for the particular case of $p_{trans}$=30 MeV/fm$^{3}$ cannot increase more than 275 MeV/fm$^{3}$ since it does not satisfy the radius constraint of the structure of 1.6 M$_{\odot}$ NS. The lower bound on the maximum mass obtained by GW190814 \cite{Mostt} also restricts our results at higher values of $\Delta\varepsilon$. Taking into account the results obtained for category III, one can notice that the increase in the value of {transition} pressure from 31 MeV/fm$^{3}$ to 60 MeV/fm$^{3}$ yields the EoS model, {in which} all the constraints can be met except the lower bound on the maximum mass obtained by GW190814. Thus, by lowering the amount of $\Delta\varepsilon$ from 350 MeV/fm$^{3}$ to 300 MeV/fm$^{3}$, the value of maximum mass is obtained around 2.08 M$_{\odot}$, which agrees well with maximum mass limits. The desirable EoS models fulfilling all these limits are the ones with $p_{trans}$$\approx$30-100 MeV/fm$^{3}$ having $\Delta\varepsilon$$\lesssim$300 MeV/fm$^{3}$, which are also in good agreement with NICER results \cite{miller, Riely}.}
\begin{table*}[t]
 \caption{{\label{twin} The values of transition density $\rho_{t}$, maximum mass of twin branch M$_{max}$ and its corresponding central density $\rho_{c}$ for the categories displayed in Figs. 4 and 5 with  particular values of $p_{trans}$ and $\Delta\varepsilon$, which are given in units of MeV/fm$^{3}$.}}
\begin{tabular}{cccc}
   \hline
    \hline
   VLOCV (AV$_{18}$)&  $\rho_{t}$ (fm$^{-3}$)&  M$_{max}$ (M$_{\odot}$) &$\rho_{c}$  (fm$^{-3}$)\\
  \hline
  CIV, $p_{trans}$=5, $\Delta\varepsilon$=350&  0.21&2.15&0.77\\
   CIV,  $p_{trans}$=30, $\Delta\varepsilon$=250    &0.33& 2.18 &0.90\\
    CIV,  $p_{trans}$=30, $\Delta\varepsilon$=268    &0.33& 2.15 &0.92\\
  CIII,  $p_{trans}$=31, $\Delta\varepsilon$=350  &0.33&2.004&0.98\\
   CIII,  $p_{trans}$=60, $\Delta\varepsilon$=350 &0.38&2.004&0.99\\
  {CIII,} {$p_{trans}$=60,} {$\Delta\varepsilon$=300} &{0.38}&{2.08}&{0.96}\\
   CIII,  $p_{trans}$=100, $\Delta\varepsilon$=300 &0.42&2.15 &0.90\\
   CII,  $p_{trans}$=112, $\Delta\varepsilon$=700 &0.43&1.72 &1.03\\
   CII,  $p_{trans}$=120, $\Delta\varepsilon$=470&0.44&1.97&0.90\\
 LOCV ({AV$_{18}$+}TNI)\\
  CIV, $p_{trans}$=4, $\Delta\varepsilon$=390   &0.19&2.1&0.77\\
   CIV, $p_{trans}$=25, $\Delta\varepsilon$=250   &0.36&2.11&0.97\\
     CIII, $p_{trans}$=41, $\Delta\varepsilon$=269   &0.42&2.003&1.08\\
    \hline
     \hline

   \end{tabular}
   \end{table*}

{It should be noted that due to the softness of the LOCV ({AV$_{18}$+}TNI) EoS compared to the VLOCV model, the instability region disconnected from the hadronic branch has been only provided for two values of $p_{trans}$, 40 and 41 MeV/fm$^{3}$, in the case of category III as displayed in the second part of Table \ref{twinn}. Moreover, {high} values of $\Delta\varepsilon$ have changed for both categories of this EoS compared to the VLOCV approach. Figure \ref{pM}, which is the mass-radius diagram of LOCV (AV$_{18}$+TNI) with the possible categories of III and IV, depicts that all the hybrid stars generated by the selection of particular values of $p_{trans}$ and $\Delta\varepsilon$ do not yield appropriate results in accord with the left-hand band (GW170817 excluded: Bauswein \textit{et al.} \cite{Bauswein}). However, all the stellar configurations do satisfy the right-hand band (GW170817 excluded: Annala \textit{et al.} \cite{Annala}) {and also the maximum mass constraint (GW170817 excluded: Rezzolla \textit{et al.} \cite{Rezzollaa}). In addition, it is easily seen that the EoS models related to the category IV meet the lower bound on the maximum mass (GW190814: Most \textit{et al.} \cite{Mostt}).}

Our obtained results for three quantities of transition density $\rho_{t}$, the maximum mass of hybrid star configurations M$_{max}$ together with its central densities $\rho_{c}$ are displayed in Table III for categories presented in Figs. \ref{WM} and \ref{pM}. $\rho_{c}$ is expected to be lower in the case of fast-rotating stars. In addition, it might serve as a guide for comparison with the densities reached in the compressed baryonic matter by heavy-ion collisions. {Furthermore, a combined analysis of GW170817 and GW190425 results in a range of densities of the neutron star core in between three to six times nuclear density, i.e., 0.48 - 0.96 fm$^{-3}$ \cite{1908425}. From this table, we can see that the compact star central density values for our models are compatible with this estimation.}}

\section{TIDAL DEFORMABILITY}
Since GW170817 has established limits on the tidal deformability of BNSMs during the inspiral phase, the EoS of these compact stars can be restricted in addition to the maximum mass constraint above 2 $M_{\odot}$. The tidal deformation of BNSs, which occurs in the early stages of the inspiral, is defined by \cite{T}
\begin{eqnarray}
\lambda=-\frac{Q_{ij}}{\varepsilon_{ij}}.
\end{eqnarray}
Indeed, the tidal deformability $\lambda$ is the ratio of the induced quadruple moment to the tidal field of its companion. It can also be related to the tidal Love number of $l=2$ through the equation below
\begin{eqnarray}
\lambda=\frac{2}{3}k_{2}R^{5}.
\end{eqnarray}
In addition, the dimensionless tidal deformability $\Lambda$ is defined by $k_{2}$ as
\begin{eqnarray}
\Lambda=\frac{2}{3}\frac{k_{2}}{\beta^{5}},
\end{eqnarray}
where $\beta=\frac{M}{R}$ is the compactness of the star. The tidal deformability is calculated along with the TOV equations in order to get the structural properties. In our calculations, we constrain our EoSs by using the precisely measured chirp mass {$\mathcal{M}=1.186 M_{\odot}$} in the inspiral phase of GW170817 \cite{abbot2019}. {Within} this work we investigate the low spin prior case for every EoS.
\begin{figure}[!ht]
\centerline{\includegraphics[scale=0.6]{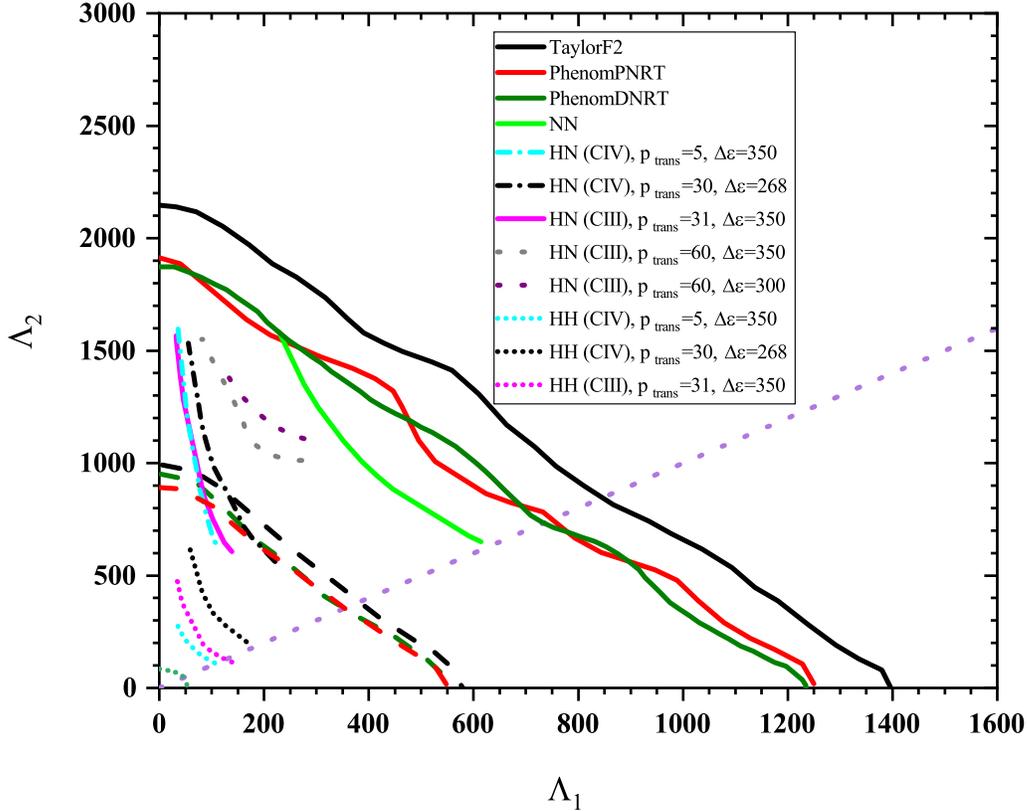}}
\caption{\small {The relation between tidal deformabilities of the components of binary system considering strong first order phase transition for the VLOCV (AV$_{18}$) model. {{The values of $p_{trans}$ and $\Delta\varepsilon$ are given in units of MeV/fm$^{3}$.~{Dashed (solid) lines correspond to the 50 $\%$ (90 $\%$) credible region for the TaylorF2, PhenomPNRT and PhenomDNRT waveform models \cite{abbot2019}.}}}}}\label{L1}
\end{figure}
\begin{figure}[!ht]
\centerline{\includegraphics[scale=0.6]{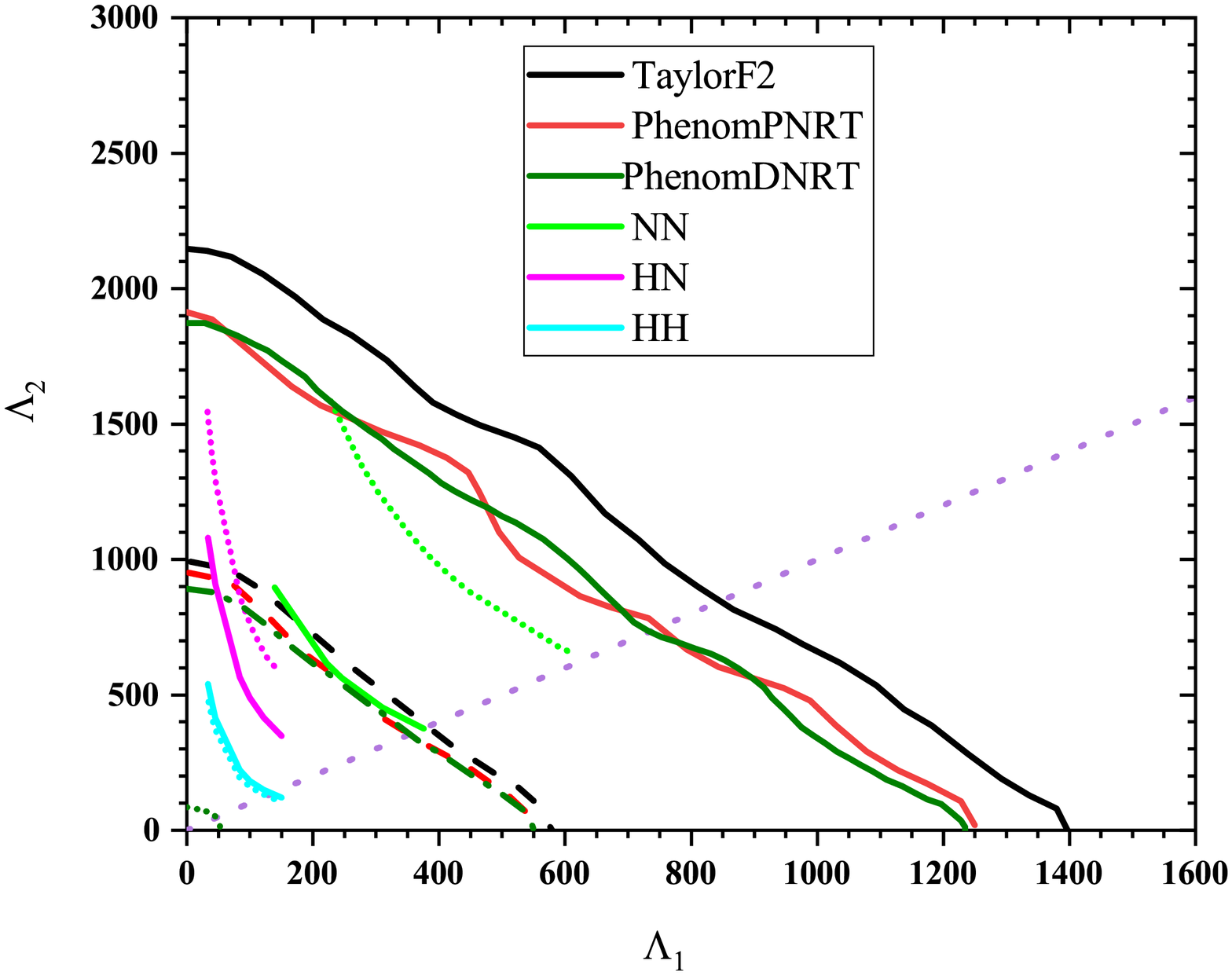}}
\caption{\small {Comparative relation between tidal deformabilities of the components of binary system for category III of both VLOCV (AV$_{18}$) and the LOCV ({AV$_{18}$+}TNI) models. {The short-dotted style refers to the results of VLOCV method, while the solid one depicts those of LOCV ({AV$_{18}$+}TNI).~{Dashed (solid) lines correspond to the 50 $\%$ (90 $\%$) credible region for the TaylorF2, PhenomPNRT and PhenomDNRT waveform models \cite{abbot2019}.}}}}\label{L3}
\end{figure}

In this section, we present our results for the tidal deformability of the components of BNSM in the form of {a} $\Lambda_{1}-\Lambda_{2}$ {diagram} to {constrain} our EoSs with GW1701817. In Fig. \ref{L1}, one can see the $\Lambda_{1}-\Lambda_{2}$ {diagram} for the equation of state related to the VLOCV model with three possible combinations of neutron star branches: neutron-neutron (NN), {hybrid-neutron} (HN), and hybrid-hybrid (HH), in which the first label refers to the massive component of the binary (M$_1$) and the second to the lighter one (M$_2$). The results are presented for categories III and IV considering particular values of $p_{trans}$ and $\Delta\varepsilon$. Note that the purple dotted line corresponds to the condition that $\Lambda_{1}=\Lambda_{2}$. The NN combination, displayed by a solid green line in this figure, is generated by the VLOCV approach without considering a phase transition. {As it is clearly seen}, this combination lies {inside} the {90$\%$ credible region which is bordered by the dashed black (dashed red/dashed dark green) line and the solid black (solid red/solid dark green) line of the probability distribution function (PDF) for the precessing waveform TaylorF2 (PhenomPNRT/PhenomDNRT)} \cite{abbot2019}.  {Moreover}, the short-dotted magenta and {cyan} lines located in the lower part of this figure are related to category III and IV, respectively. These two lines, which are representative of {low} $p_{trans}$ {(with the same value of $\Delta\varepsilon$=350 MeV/fm$^3$), lead to similar results for the tidal deformability of twin stars compared to the short-dotted black line having $p_{trans}$=30 MeV/fm$^3$ and $\Delta\varepsilon$=268 MeV/fm$^3$. These three lines located in the second branch as HH star combinations of the mentioned categories lie {within} the 50$\%$ {credible region}.}

In addition, {the solid magenta line refers to the HN star combinations generated by CIII, while the dashed-dotted {cyan} and black lines belong to CIV}. {On the one hand,} applying the {low} limit of $p_{trans}$ for both categories yields approximately identical results for $\Lambda$, while selecting greater values of $p_{trans}$, namely 60 MeV/fm$^{3}$ in the case of category III leads to {higher tidal deformability} values. On the other hand, the HN lines generated by  {low} values of $p_{trans}$ for each mentioned category are partly {inside the 50\% credible region}, whereas their major part is located {outside}. The dotted {gray} {(dark purple)} line produced by $p_{trans}$=60 MeV/fm$^{3}$ {and $\Delta\varepsilon$=350 MeV/fm$^{3}$ ($\Delta\varepsilon$=300 MeV/fm$^{3}$)} lies completely {inside the 90\%} credible levels. {We have also computed the combined dimensionless tidal deformability $\widetilde{\Lambda}$ considering the chirp mass related to the GW190425 event for these particular EoS models. In other words, by restricting our EoS to the chirp mass of 1.44 M$_{\odot}$ and the mass ratios between 0.8 to 1 leading to the total mass of 3.3 M$_{\odot}$ for the low-spin prior case, we have obtained $\widetilde{\Lambda}$ in the range of 55-100 (83-138) for the EoS model with $p_{trans}$=60 MeV/fm$^{3}$ and $\Delta\varepsilon$=350 MeV/fm$^{3}$ ($\Delta\varepsilon$=300 MeV/fm$^{3}$). Although these results agree with the value reported in \cite{1908425}, $\widetilde{\Lambda} \leq$600, this event cannot put any meaningful constraints on our EoS due to the less information on this parameter.  Unfortunately, this is also the case of the recently reported merger event GW190426, where only an estimate of the mass of the possible neutron star component of $1.5_{-0.5}^{+0.8} M_{\odot}$ is provided \cite{2020oct}.}

In our previous work \cite{sharifi}, we have computed the tidal deformability of binary neutron stars with four distinctive EoSs without phase transition, which the EoS of AV$_{18}$+TNI resulted in better outcome compared to other ones. Therefore, we have selected this EoS in order to draw a comparison between the nucleon size and TNI, which is presented in Fig. \ref{L3} for category III. It can be seen that both EoSs lead to the three possible combinations of neutron stars, which their existence in GW170817 is to some extent disputable. In this figure, the short-dotted style refers to the results of the VLOCV method, while the solid one depicts those of LOCV ({AV$_{18}$+}TNI). {One can observe that the NN combination of neutron stars is {within} the 90\% {credible region} in the case of VLOCV method {due to their large tidal deformability values}. {On the contrary, NN combinations for the LOCV (AV$_{18}$+TNI) model are marginally consistent with the 50$\%$ credible region, displaying lower tidal deformability values than the ones of VLOCV.} The HH combination for both models, located in the lower part of the figure, predicts {even} smaller tidal deformability {values} {with respect to} NN and HN combinations. Although the HN and HH combinations of the EoSs with {low} values of $p_{trans}$ for both approaches are allowed in this figure, they are ruled out by the constraint presented by Bauswein \textit{et al.} \cite{Bauswein} in the mass-radius diagrams 4 and 5.
\section{Summary and Conclusions}
In this work, we have introduced two approaches to describe the stiffness of dense matter in the core of neutron stars, the three nucleon interaction {LOCV(AV$_{18}$+TNI)} and prescription {of the excluded volume effect for nucleons} {VLOCV(AV$_{18}$)}. In all our models for the hadronic nuclear matter, we have used the two-body potential for the nuclear forces, which has recently been found to successfully describe nucleon interactions in the laboratory \cite{2020}. One of the implications of the stiff hadronic matter is the possible existence of a phase transition into deconfined quark matter in order to avoid a causality breach of the EoS. {It may happen that a hadron-quark mixed phase is unlikely to be stable for a reasonable value of surface tension \cite{Rajagopal, Neumann}. As a result, the situation is similar to the Maxwell construction case, in which two pure phases are directly in contact with each other. Therefore, the phase transition in our work is obtained by the sharp hadron to quark matter phase transition (Maxwell construction).} Thus, if the EoS features a strong first order phase transition, it will result in the mass twin scenario which brings the possibility to probe a critical end point in the QCD phase diagram.

To model the quark matter phase, we have introduced the simple constant speed of sound description and have taken the extreme limit of $c_{QM}$ equal to the speed of light. Under these assumptions, we were able to produce EoS models characterized by the critical values at the phase transition. Consequently, when solving the relativistic equations for the compact star structure, we were able to classify the resulting neutron star properties, namely the topology of sequences in the mass-radius diagram, following the convention introduced in \cite{Erik1}. We have obtained limits for these phase transition parameters in each category, which indeed satisfies the definitions according to the stellar mass. Furthermore, we have computed the tidal deformabilities of our models in order to be able to compare with the estimations derived from the compact star merger GW170817, presented in a $\Lambda_1$-$\Lambda_2$ diagram. {The combined dimensionless tidal deformability has also been computed for particular cases of category III related to the VLOCV approach in order to check the validity of our results with the GW190425 event. It turns out that this event cannot restrict our EoS models since there is little information about it.} For both models, the merger of neutron stars (NN) is possible in the light of the GW170817 event, {while neither of the other two combinations (HN and HH) is allowed for the LOCV (AV$_{18}$+TNI) model considering the Bauswein constraint \cite{Bauswein} in the mass-radius diagrams. Moreover, the HN and HH combinations of the VLOCV EoS with {low} values of $p_{trans}$ (having $\Delta\varepsilon\gtrsim$300 MeV/fm$^{3}$) for both categories of III and IV are also ruled out by this constraint.}{~It is important to note that if the limits for the maximum compact star mass are lifted, namely by new, compelling observations of massive compact stars above 2.5M$_{\odot}$ for instance of the type of the GW190814 event but undeniably featuring a compact star, then the space of parameters of our EoS models will also increase.}
\\{In this paper, we have also considered the measurement of the object PSR J0030+0451 by NICER, which allowed us to assess the validity of our results.} {It is found out that depending on the parameter values,  {transition} pressure$\approx$30-100 MeV/fm$^{3}$, and the energy density discontinuity $\Delta\varepsilon$$\lesssim$300 MeV/fm$^{3}$,} both CIII and CIV of the VLOCV model can satisfy all the observational constraints, however, CIV has more limitations. It is also deduced that in the case of the LOCV ({AV$_{18}$+}TNI) approach, neither CIII nor CIV can meet all the mentioned limits.} An astrophysical application of a category CIII EoS based on multiple polytropes can be found in \cite{alvarez7} {whereas for the CSS EoS is found in \cite{Alvarez-Castillo:2020nkp}.}

 {As all models, our EoS model has its own pros and cons.  The number of free parameters in the high-density region may be one of the limitations, as well as the only free parameter we consider in the hadronic part of our model: the radius of nucleon at equilibrium.  However, all these parameters finally lead to a space of parameter that fulfills most of the current observational constraints.
Regarding the improvements of our model, we can modify the neutron star matter and also the EoS model for the high-density region substituting the CSS parametrization model in order to be in accordance with future observations.}

\acknowledgements {We would like to appreciate the University of Zanjan Research Council. This work has been supported financially by the Research Institute for Astronomy and Astrophysics of Maragha (RIAAM) under research project No. 1/6025-16.} {D. A-C. acknowledges support from the Bogoliubov-Infeld program for collaboration between JINR and Polish Institutions as well as from the COST actions No. CA15213 (THOR) and No. CA16214 (PHAROS).}


\begin{thebibliography}{0}
\expandafter\ifx\csname natexlab\endcsname\relax\def\natexlab#1{#1}\fi
\expandafter\ifx\csname bibnamefont\endcsname\relax
  \def\bibnamefont#1{#1}\fi
\expandafter\ifx\csname bibfnamefont\endcsname\relax
  \def\bibfnamefont#1{#1}\fi
\expandafter\ifx\csname citenamefont\endcsname\relax
  \def\citenamefont#1{#1}\fi
\expandafter\ifx\csname url\endcsname\relax
  \def\url#1{\texttt{#1}}\fi
\expandafter\ifx\csname urlprefix\endcsname\relax\def\urlprefix{URL }\fi
\providecommand{\bibinfo}[2]{#2}
\providecommand{\eprint}[2][]{\url{#2}}

\end{thebibliography}


\begin{thebibliography}{}
\bibitem{Rischke} D. H. Rischke, M. I. Gorenstein, H. Stoecker, and W. Greiner, Z. Phys. C {\bf 51}, 485 (1991).
\bibitem{costa} R. Costa, A. J. Santiago, H. Rodrigues, and J. Sa Borges, Commun. Theor. Phys. {\bf46}, 1052 (2006).
\bibitem{IJME} J. Ro{\.z}ynek, Int. J. Mod. Phys. E {\bf 27}, 1850030 (2018).
{\bibitem{MIT bag model} K. Johnson,  Acta Phys. Pol. B {\bf 6}, 865 (1975).}
\bibitem{PRC91} V. Vovchenko, D. V. Anchishkin, and M. I. Gorenstein, Phys. Rev. C {\bf 91}, 064314 (2015).
{\bibitem{Typel} S. Typel, Eur. Phys. J. A {\bf 52}, 16 (2016).}
{\bibitem{ben} S. Beni\'{c}, B. Blaschke, D. Alvarez-Castillo, T. Fischer, and S. Typel, Astron. Astrophys. {\bf 577}, A40 (2015).}
{\bibitem{Cierniak:2020eyh}
M.~Cierniak and D.~Blaschke,
Eur. Phys. J. Special Topics \textbf{229}, 3663 (2020).}
{\bibitem{122} E. R. Most, L. J. Papenfort, V. Dexheimer, M. Hanauske, S. Schramm, H. St\"{o}cker, and L. Rezzolla, Phys. Rev. Lett. {\bf 122}, 061101 (2019).
\bibitem{2} E. Witten, Phys. Rev. D {\bf 30}, 272 (1984).
\bibitem{3} F. Weber, Prog. Part. Nucl. Phys. {\bf 54}, 193 (2005).
\bibitem{4} J. M. Lattimer and M. Prakash, Phys. Rep. {\bf 442}, 109 (2007).
\bibitem{5} A. Zacchi, R. Stiele, and J. Schaffner-Bielich, Phys. Rev. D {\bf 92}, 045022 (2015).
\bibitem{Erik2}  J.-E. Christian, A. Zacchi, and J. Schaffner-Bielich, Phys. Rev. D {\bf 99}, 023009 (2019).
\bibitem{6} Z. Fodor and S. D.  Katz, J. High Energy Phys. {\bf 03}, (2002) 014.}
{\bibitem{seidov} Z. F. Seidov, Sov. Astron. {\bf 15}, 347 (1971).}
{\bibitem{alvarez} D. Alvarez-Castillo, A. Ayriyan, S. Beni\'{c}, D. Blaschke, H. Grigorian, and S. Typel, Eur. Phys. J. A {\bf 52}, 69 (2016).}
{\bibitem{alvarez2} A. Ayriyan, D. Alvarez-Castillo, D. Blaschke, and H. Grigorian, Universe {\bf 5}, 61 (2019).}
{\bibitem{alvarez3} D. Alvarez-Castillo, S. Beni\'{c}, D. Blaschke, S. Han, and S. Typel, Eur. Phys. J. A {\bf 52}, 232 (2016).}
{\bibitem{alvarez4} V. Paschalidis, K. Yagi, D. Alvarez-Castillo, D. B. Blaschke, and A. Sedrakian, Phys. Rev. D {\bf 97}, 084038 (2018).}
{\bibitem{alvarez5} D. E. Alvarez-Castillo, D. B. Blaschke, A. G. Grunfeld, and V. P. Pagura, Phys. Rev. D {\bf 99}, 063010 (2019).}
{\bibitem{Montana}  G. Montana, L. Tol\'{o}s, M. Hanauske, and L. Rezzolla, Phys. Rev. D {\bf 99}, 103009 (2019).}
\bibitem{chris} J. E. Christian and J. Schaffner-Bielich, Astrophys. J. Lett. {\bf 894}, L8 (2020).
\bibitem{alvarez6} D. Blaschke, D. E. Alvarez-Castillo, A. Ayriyan, H. Grigorian, N. K. Largani, and F. Weber, (World Scientific, Singapore, 2020), chap. 7, pp. 207-256.

\bibitem{Ayriyan:2021prr}
A.~Ayriyan, D.~Blaschke, A.~G.~Grunfeld, D.~Alvarez-Castillo, H.~Grigorian, and V.~Abgaryan,
arXiv:2102.13485.
{\bibitem{124} D. Blaschke and D. Alvarez-Castillo, Eur. Phys. J. A {\bf 56}, 124 (2020).}
{\bibitem{Zhaoo} T. Zhao and J. M. Lattimer, Phys. Rev. D {\bf 102}, 023021 (2020).}
{\bibitem{Biswas} B. Biswas, P. Char, R. Nandi, and S. Bose, arXiv:2008.01582.}
{
\bibitem{Tan:2020ics}
H.~Tan, J.~Noronha-Hostler, and N.~Yunes,
Phys. Rev. Lett. \textbf{125}, 261104 (2020).}
{\bibitem{Asadi} Z. A. Aghbolaghi and M. Bigdeli, J. Phys. G {\bf 45}, 065101 (2018);} Z. A. Aghbolaghi and M. Bigdeli, Eur. Phys. J. Plus {\bf 134}, 430 (2019).
\bibitem{Erik1} J.-E. Christian, A. Zacchi, and J. Schaffner-Bielich, Eur. Phys. J. A {\bf 54}, 28 (2018).
{\bibitem{1908425} B. P. Abbott \textit{et al.}, Astrophys. J. Lett. {\bf 892}, L3 (2020).}
{\bibitem{reanalysis} T. Narikawa, N. Uchikata, K. Kawaguchi, K. Kiuchi, K. Kyutoku, M. Shibata, and  H. Tagoshi, Phys. Rev. Research {\bf 2}, 043039 (2020).}
\bibitem{clark} J. W. Clark and N. C. Chao, {Lett. Nuovo Cimento} {\bf 2}, 185 (1969).
\bibitem{OBI3} J. C. Owen, R. F. Bishop, and J. M. Irvine, {Nucl. Phys.}  A{\bf277}, 45 (1977).
\bibitem{wiringa} R. B. Wiringa, V. G. J. Stoks, and R. Schiavilla, Phys. Rev. C {\bf 51}, 38 (1995).
{\bibitem{Lagaris} I. E. Lagaris and V. R. Pandharipande, Nucl. Phys. A{\bf 359}, 349 (1981).}
\bibitem{BM98} G. H. Bordbar and M. Modarres, Phys. Rev. C {\bf57}, 714 (1998).
{\bibitem{quark meson} K. Saito and A. W. Thomas, Phys. Lett. B {\bf 327}, 9 (1994).}
\bibitem{Gue} G. Hua, J. Phys. G {\bf 25}, 1701 (1999).
\bibitem{guch} P. A. Guichon, Phys. Lett. B {\bf 200}, 235 (1988).
\bibitem{nuclear} J. Ro{\.z}ynek, J. Phys. G {\bf 42}, 045109 (2015).
\bibitem{Elyasi}  M. Bigdeli and S. Elyasi, Eur. Phys. J. A {\bf51}, 38 (2015).
\bibitem{Alford} M. G. Alford, G. F. Burgio, S. Han, G. Taranto, and D. Zappala, Phys. Rev. D {\bf 92}, 083002 (2015).
\bibitem{glendening} N. K. Glendenning and C. Kettner, Astron. Astrophys. {\bf 353}, L9 (2000).
\bibitem{sch} J.  Schaffner-Bielich,  M.  Hanauske,  H.  St\"{o}cker, and W. Greiner, Phys. Rev. Lett. {\bf 89}, 171101 (2002).
\bibitem{zacchi} A. Zacchi, L. Tolos, and J. Schaffner-Bielich, Phys. Rev. D {\bf 95}, 103008 (2017).
\bibitem{Tolman} R. C. Tolman, Phys. Rev. {\bf 55}, 364 (1939).
\bibitem{Oppenheimer} J. Oppenheimer and G. Volkoff, Phys. Rev. {\bf 55}, 374 (1939).
{\bibitem{Rezzollaa} L. Rezzolla, E. R. Most Elias, and R. Weih Lukas, Astrophys. J. Lett. {\bf 852}, L25 (2018).}
{\bibitem{Bauswein} A. Bauswein, O. Just, H. T. Janka, and N. Stergioulas, Astrophys. J. Lett. {\bf 850}, L34 (2017).}
{\bibitem{Annala} E. Annala, T. Gorda, A. Kurkela, and A. Vuorinen, Phys. Rev. Lett. {\bf 120}, 172703 (2018).}
{\bibitem{190814} R. Abbott, \textit{et al.}, Astrophys. J. Lett. {\bf 896}, L44 (2020).}
{\bibitem{Mostt} E. R. Most, L. J. Papenfort, L. R. Weih, and L. Rezzolla, {Mon. Not. R. Astron. Soc. {\bf 499}, L82 (2020).}}
\bibitem{miller} M. C. Miller \textit{et al.}, Astrophys. J. Lett. {\bf {887}}, L24 (2019).
\bibitem{Riely} T. E. Riley \textit{et al.}, Astrophys. J. Lett. {\bf {887}}, L21 (2019).
\bibitem{T} T. Hinderer, Astrophys. J. {\bf 677}, 1216 (2008).
\bibitem{abbot2019} B. P. Abbott \textit{et al.}, Phys. Rev. X {\bf 9}, 011001 (2019).
\bibitem{2020oct} R. Abbott \textit{et al.}, arXiv:2010.14527.
\bibitem{sharifi} Z. Sharifi and M. Bigdeli, J. Phys. G {\bf 46}, 125203 {(2019)}.
\bibitem{2020} A. Schmidt \textit{et al.}, Nature (London) {\bf {578}}, 540 (2020).
\bibitem{Rajagopal} M. G. Alford, K. Rajagopal, S. Reddy, and F. Wilczek, Phys. Rev. D {\bf 64}, 074017 (2001).
\bibitem{Neumann} F. Neumann, M. Buballa, and M. Oertel, Nucl. Phys. A{\bf 714}, 481 (2003)
\bibitem{alvarez7} D. E. Alvarez-Castillo, J. Antoniadis, A. Ayriyan, D.
Blaschke, V. Danchev, H. Grigorian, N. K. Largani, and
F. Weber, Astron. Nachr. {\bf 340}, 878 (2019).
{
\bibitem{Alvarez-Castillo:2020nkp}
D.~E.~Alvarez-Castillo,
Astron. Nachr. \textbf{342}, 234 (2021).}

\end{thebibliography}
\end{document}